\documentclass[a4paper,preprintnumbers,showpacs,twocolumn,superscriptaddress,nofootinbib,amsmath,amssymb]{revtex4-1}
\usepackage{graphics}
\usepackage{color,hyperref}
\usepackage{times}
\usepackage{mathrsfs}
\hypersetup{
  colorlinks=true,
  linkcolor=blue,
  citecolor=magenta,
  filecolor=magenta,
  urlcolor=blue
}

\def\be{\begin{equation}}
\def\ee{\end{equation}}
\def\bea{\begin{eqnarray}}
\def\eea{\end{eqnarray}}

\usepackage{enumitem}
\usepackage{graphicx,subfigure}% Include figure files
\usepackage{dcolumn}% Align table columns on decimal point
\usepackage{bm}% bold math
\usepackage{color}

\begin{document}

\title{On causal structure of $4D$-Einstein-Gauss-Bonnet black hole}

\author{Naresh Dadhich}
\email{nkd@iucaa.in}
\affiliation{Inter-University Centre for Astronomy and Astrophysics, Post Bag 4, Ganeshkhind, Pune 411 007, India}

\begin{abstract}
The recently proposed effective equation of motion for the $4D$- Einstein-Gauss-Bonnet gravity admits a static black hole solution that has, like the Rissner-Nordstr\"{o}m charged black hole,  two horizons instead of one for the Schwarzschild black hole. This means that the central singularity is timelike instead of spacelike. It should though be noted that in $D\geq5$, the solution always admits only one horizon likee the Schwarzshild solution. In the equation defining the horizon, the rescaled Gauss-Bonnet coupling constant appears as a new 'gravitational charge' with a repulsive effect to cause in addition to event horizon a Cauchy horizon. Thus it radically alters the causal structure of the black hole.

\end{abstract}

\maketitle

%\section{Introduction}
It is well known that the Lovelock theory \cite{lovelock}, whose action is a homogeneous polynomial in Riemann curvature, is the most natural higher dimensional generalization of the Einstein gravity --- general relativity (GR). It has the remarkable property that despite the action being polynomial in Riemann curvature, yet the equation of motion remains second order. However the higher order terms in the action make non-zero contribution in the equation only in dimension $D>2N$ where $N$ is the degree of curvature polynomial in the action. GR is linear  while Gauss-Bonnet (GB) is quadratic order Lovelock and so on. Thus Lovelock is quintessentially a higher dimensional natural generalization of GR. \\

It is however possible to make higher order terms contribute in the equation in $4D$ by dilaton coupling --- a scalar field coupled to higher order term in the action, see for instance \cite{kanti}. Recently a new proposal has been made \cite{glavan} wherein GB term is made to contribute in $4D$ without dilaton coupling. In that the GB coupling is scaled as $\alpha \rightarrow \alpha/(D-4)$ and thereby cancelling out $(D-4)$ factor in the equation, and then taking the limit $D\to4$. This results into an effective equation in $4D$ which is in fact the Einstein-Gauss-Bonnet (EGB) equation written for $D=4$ \footnote{The equation is non-vacuous only in $D>4$ because of the multiplicative factor $(D-4)$, which has now been cancelled out by rescaling of $\alpha$.}. Then it could be solved in spacetime  with some specific symmetries for different situations, black holes and cosmology. Firstly this way of taking limit is rather contentious and most importantly, there is no corresponding $4D$-action for the equation. Not withstanding all this, it has instantly caught up like a wild fire as is evidenced by the runaway  activity \cite{fire} which is still going stronger by the day as there continues to be a steady flow of papers on the arxiv. \\

On the other hand there are some serious questions being posed on the overall acceptability of the limiting process, validity of the equation in $4D$ as well as absence of proper action and a consistent theory in $4D$ \cite{ai, gurses, mann, clifton, mukohyama, mahapatra, kurt, cano, gaurav}. In particular It is fair to say that the jury is out on this issue, and we have to wait for some time before the air is cleared. \\

In this letter we wish to take up the issue of the static black hole solution of the new proposed $4D$-EGB equation that admits two horizons instead of the usual one for the Schwarzschild solution. On the other hand the EGB equation has only one horizon like the Schwarzschild in $D\geq5$ which is the natural rightful playground   for it. In transition to $4D$, it acquires an extra Cauchy horizon which indicates presence of a 'new charge'. How does that arise physically and how do we understand it ? These are the most pertinent and critical questions. \\

Let's begin by recalling the $4D$-EGB static black hole metric \cite{glavan} as given by
\begin{equation}
ds^2 = - f(r)dt^2 + \frac{dr^2}{f(r)} + r^2 d\Omega^2
\end{equation}
where
\begin{equation}
f(r) = 1+\frac{r^2}{2\alpha}\left(1\pm\sqrt{1+\frac{8\alpha M}{r^3}}\right).
\end{equation}
In the usual notation $d\Omega^2$ is the metric on unit $2$-sphere and $\alpha$ and $M$ are the rescaled GB coupling constant and mass of isolated body respectively. We have also set $G=c=1$. The negative sign is chosen in the above solution for gravity being attractive. \\

The black hole horizon is given by $f(r)=0$ which solves to give horizons as
\begin{equation}
r_{h\pm} = M(1 \pm \sqrt{1 - \alpha/M^2}) .
\end{equation}
Thus black hole has two horizons, unless of course $\alpha<0$ \footnote{Since gravity is universal and hence always attractive \cite{dadhich1}, its coupling constant should always have the same sign for all Lovelock orders $N$. Therefore $\alpha$ cannot be negative. Since it is a dimensionful, its dimension will however  depend on order $N$ and, consequently or otherwise also on spacetime dimension.}, with the condition $M^2\geq \alpha$. The two merge into one-another for $M^2=\alpha$, defining the extremality condition. For $\alpha=0, M^2$, the event horizon is respectively $r_h = 2M, M$. This is exactly like the Reissner-Nordstr\"{o}m metric where $\alpha$ has replaced charge $Q^2$. That means it has acquired gravitational charge character which is rather very strange and queer --- a coupling constant being a charge! It produces repulsive effect exactly like the Maxwell charge for charged black hole so as to create a Cauchy horizon. In a different context the same repulsive effect has also been found \cite{seyed}.  \\

Note that we have the famous Boulware-Deser EGB black hole \cite{boulware} solution which could be written for any $D>4$ with
\begin{equation}
f(r) = 1 + \frac{r^2}{2\alpha}\left(1\pm\sqrt{1+\frac{8\alpha M}{r^{D-1}}}\right)
\end{equation}
where $\alpha$ is normalized for a $D$ dependent numerical factor. It is this solution which is written in Eq. (2) above for $D=4$ as the solution of $4D$-EGB equation. The horizon equation $f(r)=0$ would then take the form for a generic dimension $D$,
\begin{equation}
r^{D-3} + \alpha r^{D-5} - 2M = 0.
\end{equation}

Clearly this equation admits only one positive root for $D\geq5$ while two positive roots for $D=4$. That means EGB black hole has only one event horizon for $D\geq5$ while it has two --- both event and Cauchy horizons for $D=4$. Further it is clear from Eq. (4) that the metric is regular at $r=0$ for $D=4, 5$ \cite{dadhich2} but curvatures diverge with a lesser power of $r$ --- singularity is weakened. This is the GB effect resulting in weakening of gravity.\\ 

In transition from five to four dimension, spacetime structure has radically changed. In the former the central singularity is spacelike as for the Schwarzschild black hole in any dimension $D\geq4$ in GR. So is also the case for EGB black hole in $D\geq5$. This means static black hole in Einstein gravity for $D\geq4$ --- $D$-dimensional Schwarzschild and in EGB gravity $D$-dimensional Boulware-Deser for $D\geq5$ share the same causal structure \cite{torii} having spacelike central singularity. In contrast $4D$-EGB black hole has radically different causal structure with central singularity being timelike. It shares the structure instead of the Schwarzschild with Reissner-Nordstr\"{o}m charged black hole. \\

The key questions that arise are: Since no additional matter field has been introduced, what is it that causes this radical change in spacetime structure? The Cauchy horizon is always caused by some new 'charge' like Maxwell charge or rotation which has opposing repulsive contribution to mass. In here nothing of that sort has happened, and it has been left to the GB coupling $\alpha$ to do the job \footnote{It should be noted that $\alpha$ appears as a 'charge' only in $D=4$ while for in all $D\geq5$, the black hole admits only one horizon. It is purely a four-dimensional feature.}. A coupling constant should not serve as gravitational charge,  it is the measure of field's linkage to matter. \\

One of the motivations for higher curvature theories, in particular Lovelock, is that when one probes high energy strong field limit of GR, it is pertinent to include higher powers of Riemann curvature in action \cite{dadhich2}. Yet if the equation of motion should not change its second order character, the Lovelock generalization is then unique and it plays out only in higher dimensions. This is how higher dimensions, $D>4$, are innately tied to the Lovelock gravity. Of course one would like to bring down higher curvature effects that rule in higher dimensions to four dimensions to confront them against observations. That has always been accomplished, as mentioned earlier, through dilaton coupling --- a scalar field coupled to higher order Lovelock term in action. What is envisaged is that the effect in four dimension should appear as correction to GR rather than a radical departure from it. Unless of course, there is some new property or feature of gravity has been unravelled which has so far remained unprobed. \\

From this standpoint what ensues in $4D$-EGB is entirely different and affront. It radically alters the causal structure by letting the GB coupling behave like a 'charge'. This is rather queer and strange, and hard to understand. We do however have a similar situation in the brane-world gravity model \cite{randall}. There a black hole on the $3$-brane does acquire a new charge --- the Weyl charge that arises from the Weyl curvature of bulk spacetime  \cite{dmpr}. The metric is exactly that of the Reissner-Nordstr\"{o}m charged black hole. Then the sign of the Weyl charge, analogue of $Q^2$ is taken as negative so as to have only one horizon. It was envisaged that modification ensuing from bulk should only produce 'correction' to GR black hole without any significant modification to spacetime structure. The Weyl charge is sourced by Weyl curvature of bulk spacetime. It is through the Weyl charge that higher dimensional bulk geometry manifests in four dimension as a correction to GR. Like the Maxwell charge for charged Reissner-Nordstr\"{o}m black hole, the Weyl charge is the ADM charge for a black hole on the brane. \\

Despite there being imprint of two charges on the horizon, yet the spacetime asymptotically goes over to the Schwarzschild solution with mass alone as the parameter. At infinity it is only mass, which is the ADM charge. Since no matter field of any kind has been added, there is no charge that can come into play. It is the GB gravitational coupling constant that manifests as charge only on the horizon. Its role as 'charge' disappears at large $r$. Conceptually this is the most discomforting aspect of this proposal. Gravity resides in spacetime geometry while gravitational charge in matter fields. Here there is no identifiable matter field that could be responsible for creation of the Cauchy horizon. \\

The authors of the proposal \cite{glavan} perceive that the Gauss-Bonnet equation is a classical modification to GR and not a one loop correction, and hence it is on the same footing as GR. This is very well for $D>4$ in its rightful playground. The question is, should its effect obtained in whatever way in four dimension be a correction to GR solutions or a radical change? The proposal implies the latter as indicated by the two horizons of a static black hole. They envisage that outer horizon is the event horizon for black hole while the inner one (which is the Cauchy horizon for black hole)is the event horizon for white hole. Gravitational collapse comes to a halt in between the two horizons, and collapse turns into expansion. That is how a white hole emerges avoiding central singularity. The metric is regular\footnote{It may be noted that the metric is also regular in $D=5$, the Boulware-Deser black hole \cite{boulware}. } at $r=0$ but Riemann curvature however diverges with weaker fall off as $r^{-3/2}$ indicating weakening of the singularity. \\

This is all very fine provided we are able to identify some physical property or principle that is responsible for turning attraction into repulsion, and black hole into white hole. In the absence of new matter field or quantum effects, what is it that so radically alters the nature of gravity. This is the real question, and that is what is required to be addressed and answered. That is all that we wish to raise here. \\

In a very recent paper \cite{ge}, it has been shown by causal structure analysis of bulk spacetime that the GB coupling is bounded from above in AdS space, $\alpha\leq0$;i.e. it is non-positive. In that case there will be only one horizon thereby recovering the Schwarzschild causal structure. But then the GB gravity would be repulsive in stark opposition to the Einstein gravity. This is certainly not acceptable. Gravity should have the same character at all Lovelock orders, it cannot suddenly turn repulsive for $N=2$. If true, this raises a clear and sharp question for overall tenability of the $4D$-EGB proposal. \\

Finally it is the rescaling of the GB coupling that has brought the otherwise vacuous equation in $D=4$ onto four dimension, and it is that which has turned a black hole into a white hole! As emphasized earlier, in all $D\geq5$, the solution always admits only one horizon. This strange intriguing feature is purely due to the questionable limiting procedure. In addition of course there is no valid action and a consistent theory in four dimension. \\

\section*{Acknowledgement} I thank Sumanta Chakraborty for useful discussion.

\end{document}